# Construction of Kondo Chains by Engineering Porphyrin π-Radicals on Au(111)

*Yan Zhao[1,†], Kaiyue Jiang[2,†], Peng-Yi Liu[3,†], Jie Li[1,†], Ruoning Li[4], Xin Li[1], Xinchen Fang[1], Anjing Zhao[4], Yutong Zhu[1], Hongxiang Xu[1], Ting Chen[4,*], Dong Wang[4], Xiaodong Zhuang[2,*], Shimin Hou[1], Kai Wu[5], Song Gao[6], Qing-Feng Sun[3,7,*], Yajie Zhang[1,*] and Yongfeng Wang[1,*]*

*[†]These authors contributed equally to the work.*

*[*]Corresponding Authors: yongfengwang@pku.edu.cn; yjzhang11@pku.edu.cn; sunqf@pku.edu.cn; zhuang@sjtu.edu.cn; chenting@iccas.ac.cn.*

[1]Research Center for Carbon-based Electronics and Key Laboratory for the Physics and Chemistry of Nanodevices, School of Electronics, Peking University, Beijing 100871, China.

[2]The Soft2D Lab, State Key Laboratory of Synergistic Chem-Bio Synthesis, School of Chemistry and Chemical Engineering, Shanghai Jiao Tong University, 130 Dongchuan Road, Shanghai 200240, China

[3]International Center for Quantum Materials, School of Physics, Peking University, Beijing 100871, China

[4]CAS Key Laboratory of Molecular Nanostructure and Nanotechnology, CAS Research/Education Center for Excellence in Molecular Sciences, Beijing National Laboratory for Molecular Sciences (BNLMS), Institute of Chemistry, Chinese Academy of Sciences, Beijing 100190, China

[5]BNLMS, College of Chemistry and Molecular Engineering, Peking University, Beijing 100871, China

[6]Spin-X Institute, School of Chemistry and Chemical Engineering, South China University of Technology, Guangzhou 510641, China

[7]Hefei National Laboratory, Hefei 230088, China.

**Abstract**

Quantum manipulation of molecular radical spins provides a crucial platform for exploring emergent phenomena in many-body systems. Here, we combine surface-confined synthesis with scanning tunneling microscopy (STM) tip-induced dehydrogenation to achieve atom-precise engineering of quasi-one-dimensional porphyrin-based Kondo chains (1–7 units) on Au(111). High-resolution STS measurements and low-energy effective modeling collectively demonstrate that π-radicals at each fused-porphyrin unit form Kondo singlets screened by conduction electrons. Adjacent singlets develop direct coherent coupling via quantum-state-overlap-enabled electron tunneling. Crucially, chiral symmetry in the effective model governs zero-mode distribution—present in odd-length chains yet absent in even-length chains—which dictates pronounced odd-even quantum effects in STS spectra of finite chains. Furthermore, the number of parallel porphyrin chains non-monotonically tunes the competition between the Kondo effect and spin exchange, showing opposing trends in strength and demonstrating that both wave-function overlap and the SOMO–LUMO gap collectively govern these interactions. This work simultaneously resolves the dimensional dependence of many-body correlations in confined quantum systems and pioneers approaches for quantum-critical manipulation in molecular spin architectures.

## Introduction

Research on many-body effects in strongly correlated electron systems constitutes a core scientific challenge for uncovering the microscopic mechanisms underpinning exotic quantum states of matter.[1–3] The Kondo effect, a cornerstone of such studies, exemplifies many-body entanglement between localized spins and conduction electrons. It arises when conduction electrons screen a local magnetic moment, forming a nonmagnetic singlet state.[4] However, in periodic arrays of magnetic centers, the Kondo effect extends beyond single-impurity scenarios. In such configurations, the Ruderman-Kittel-Kasuya-Yosida (RKKY) interaction competes or cooperates with Kondo screening, driving the system toward distinct ground states—paramagnetic phases, magnetic order, or even coexisting regimes where Kondo and magnetic order intertwine.[5]

Breakthroughs in low-temperature scanning tunneling microscopy (STM) provide revolutionary capabilities for real-space investigation of Kondo physics and its quantum competition with magnetic interactions.[6–8] By leveraging atomic-resolution tip manipulation, researchers artificially engineer magnetic systems—ranging from quantum dots[9] to single atoms[10,11] and low-dimensional spin arrays[12–15]—enabling systematic exploration of correlated many-body phenomena across the spectrum from local Kondo screening to long-range magnetic exchange. However, traditional *d-/f*-electron systems remain constrained by strong localization: spin-lattice coupling induced by surface-derived crystal fields generates complex magnetic anisotropy, hindering atomic-scale quantum state control and mechanistic resolution.

In contrast, π-radical organic magnet systems[16–18] offer distinct advantages. Spin degrees of freedom can be precisely engineered through molecular topology design.[19–24] The low atomic mass of carbon atoms yields extremely weak magnetic anisotropy and spin-orbit coupling, resulting in extended spin coherence times and lengths.[25–26] The delocalized nature of π-radicals facilitates efficient magnetic exchange interactions, providing an ideal, tunable platform for investigating many-body correlations. Recent advances in bottom-up surface synthesis techniques[27,28] have established organic π-radical as prototype systems for exploring quantum correlated phases. Experimentally realized structures include one-dimensional S=1 spin chains exhibiting Haldane gaps and topological edge states,[29,30] S=1/2 alternating-exchange Heisenberg model,[31] and S=1/2 Heisenberg chains demonstrating spinon continuum characteristics.[32–36]

Nevertheless, in such reported systems, one-dimensional antiferromagnetic chains formed from molecular building blocks exhibit significantly elevated spin density on carbon atoms due to their small size. This triggers strong intrachain antiferromagnetic coupling, causing the exchange interaction between adjacent units to dominate—far exceeding the chain-substrate coupling and thereby suppressing the Kondo effect. While important pioneering work has demonstrated Kondo chain behavior in inorganic atomic systems,[37,38] studies aiming to realize Kondo chains in organic systems have not yet reported the site-dependent zero-bias resonance signature that would provide direct evidence of inter-singlet coupling between neighboring Kondo singlets[39,40] Achieving authentic one-dimensional Kondo chains in organic systems requires employing larger molecules to enhance π-electron spin delocalization. This reduces spin density per carbon atom and substantially weakens intrachain antiferromagnetic interactions.

Here, we select large molecular-sized porphyrin precursor Zinc(II) 5,15-bis(4-bromo-2,6-dimethylphenyl)porphyrin (Zn(II)Por(dmp)2-2Br) as building unit. Combining on-surface reactions with STM tip manipulation techniques, we constructed fused-porphyrin organic Kondo chains on Au(111) ranging from 1 to 7 units in length. Each porphyrin unit hosts a π-radical spin screened into a Kondo singlet by the Au(111) electron reservoir. Adjacent singlets become correlated through wavefunction overlap, forming a quasi-one-dimensional S = 1/2 Kondo chain. Joint experimental and numerical results reveal that zero-energy mode distribution in open-boundary chains drives pronounced odd-even quantum effects. Odd-numbered chains exhibit robust zero-bias Kondo peaks at odd sites, while even-numbered chains develop asymmetric states split about the Fermi level. Furthermore, the number of parallel porphyrin chains (single, fused-double, and fused-triple) non-monotonically modulates the competition between the Kondo effect and spin exchange coupling, with Kondo strength decreasing as single > fused-triple > fused-double, while spin exchange follows the opposite trend. Our finding demonstrates that both wavefunction overlap and the SOMO–LUMO gap collectively govern the strength of spin exchange interactions and Kondo coupling.

**Results and Discussion**

The synthesis of fused-porphyrin chains exhibiting π-radical magnetism is challenging via wet-chemical methods due to poor solubility, strong π-π stacking interactions, and high reactivity. To overcome these obstacles, this study employs a state-of-the-art strategy based on-surface synthesis and STM atomic manipulation.[30,41] The approach utilizes the modified porphyrin precursor Zn(II)Por(dmp)$_2$-2Br (Figure 1a, M1), whose bromine-functionalized terminals promote chain extension. Figure 1a-d illustrates the strategy for constructing fused-porphyrin chains on Au(111). Initially, high-purity M1 molecules are synthesized through solution-phase methods (synthetic route in Supporting information Figure S1). Subsequently, molecules are sublimated onto Au(111) under ultrahigh vacuum, followed by thermal annealing at 433 K to initiate Ullmann coupling, forming oligomer chains of varying lengths (Figure 1a). A annealing (5 K/min ramp to 533 K) triggers methyl cyclodehydrogenation. During this process, dissociated hydrogen atoms migrate across the surface and saturate radical sites, generating *sp³*-hybridized methylene (-$CH_2$-) groups at porphyrin outer corners (Figure 1b)[30]. Ultimately, proximal chains undergo fusion, producing fully aromatic porphyrin ribbons with tunable widths (Figure S2). Non-contact atomic force microscopy (Nc-AFM, Figure S6b) image reveals that within the fused-double-porphyrin chains, dehydrogenated methyl groups of porphyrin units form regular hexagonal rings at their peripheries, with adjacent units covalently linked via three new C-C bonds. Critically, methylene species persist along non-fused segments (Figure 1c).[30,41] In addition, high-resolved STM further confirms that five-membered rings containing methylene exhibit larger spatial dimensions and sharper geometric angles versus other rings (Figure S6c).[21,42]

To precisely tailor spin distributions, as indicated by the yellow markers in Figure 1c, we positioned the STM tip over targeted *sp³*-hybridized carbon sites and applied bias voltage ≥2.6 V to selectively dissociate C–H bonds at these locations (see Supplementary information for details). This process controllably converts the designated *sp³*-hybridized carbon sites one-by-one into *sp²*-hybridized configurations, generating delocalized π-radicals at predefined positions (Figure 1d). The complete evolution from initial closed-shell fused-double-porphyrin to final antiferromagnetic spin chain is captured in Figures 1e-f, showing the sequential generation of seven S = 1/2 spins. The accompanying STM images reveal the progressive emergence of new electronic states near Fermi level (Figure 1f), while high-resolution Nc-AFM (Figure S6e) directly verifies this structural transition.

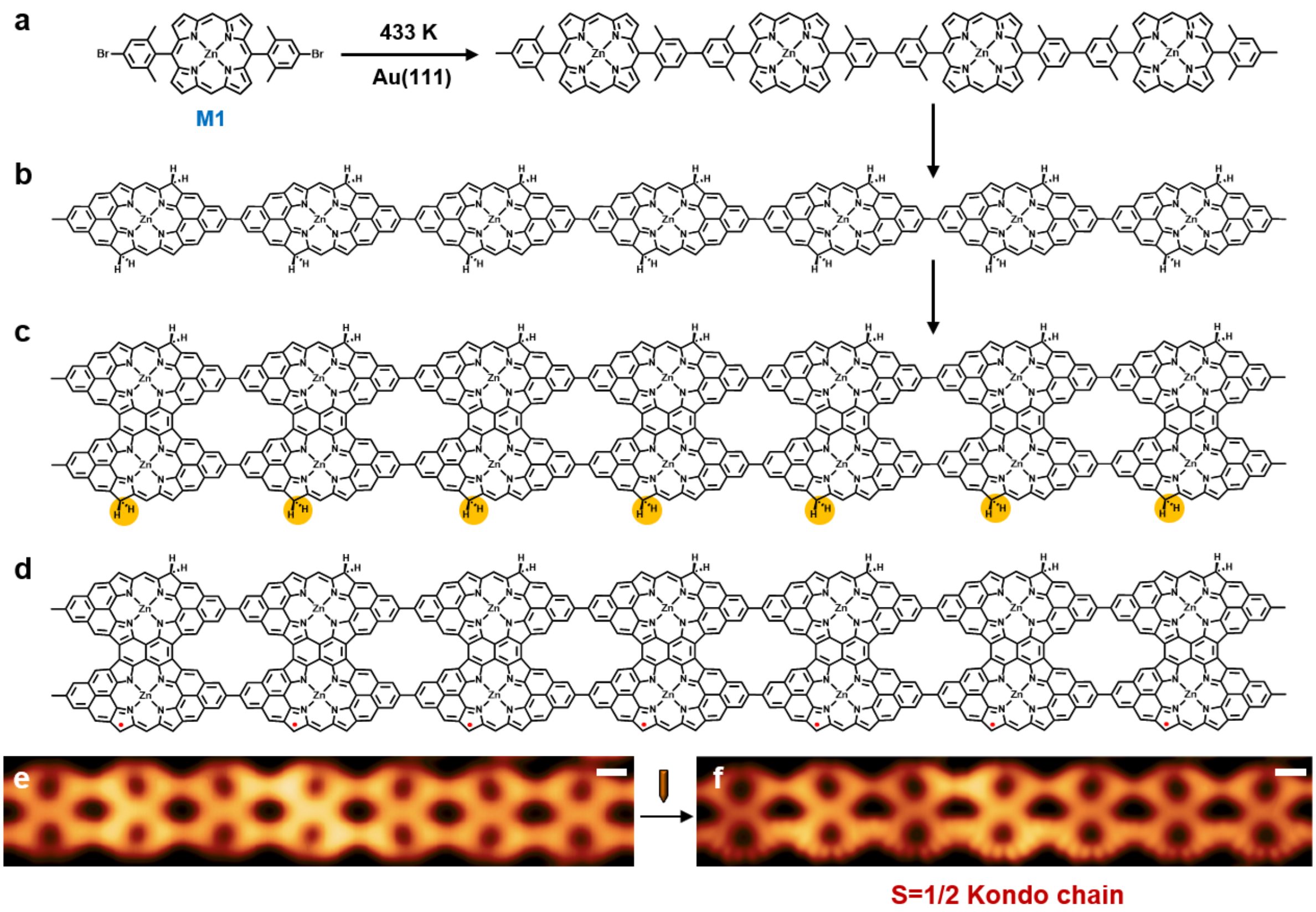

**Figure 1.** (a-c) On-surface synthesis of fused-porphyrin chains. (d) Tip manipulation process. The yellow markers in panel (c) indicate target sites for subsequent tip-induced dehydrogenation. (e-f) STM topographic images before and after tip manipulation. (e: $V$ = 0.03 V, $I$ = 1 nA, Scale bar: 0.5 nm; f: $V$ = 0.05 V, $I$ = 50 pA, Scale bar: 0.5 nm)

By constructing fused-double-porphyrin chains with site-specific radical control (Figure 2), we probe Kondo interactions (π-radical/substrate) and inter-radical magnetic coupling. When a π-radical monomer is engineered on one side of the fused-double chain, STM topological image shows density of states distribution primarily concentrated on the dehydrogenated side (Figure 2a), exhibiting excellent agreement with DFT-calculated spin density distributions (Figure 2b). Corresponding differential conductance (d$I$/d$V$) spectra feature a sharp zero-bias peak (Figure 2c)—signature of Kondo resonance from conduction electron screening of a magnetic impurity.[43] This confirms an S=1/2 ground state with relatively strong Kondo correlation to the substrate[44–47]. Fitting the Kondo resonance with a Frota lineshape[48,49] yields a half-width at half-maximum (HWHM) $\Gamma$= 1.02 meV. The energy scale of the Kondo effect is represented by the Kondo temperature $T_K$,[50] directly related to the resonance HWHM. Based on the relation $\Gamma \sim k_B T_K$, we estimate $T_K$~11.8 K.

Subsequently, we engineered a second π-radical on the same side to probe spin-spin interactions. STM imaging reveals density of states distribution of the diradical on the identical side (Figure 2d), with spatial features matching DFT-calculated spin density image (Figure 2e). Gas-phase DFT calculations (Figure S12b) indicate antiferromagnetic exchange $J_H$ in this diradical system, forming a singlet ground state (S=0) with a triplet (S=1) excitation gap of 4.2 meV. Experimentally, d*I*/d*V* spectra measured at two sites corresponding to Figure 2d show two asymmetric low-energy peaks relative to the Fermi level (Figure 2f). Fitting multiple spectra using the Ternes model[51] yields an antiferromagnetic exchange strength ranging from 3.1 to 5.8 meV (Figure S13c), which fluctuates around the gas-phase DFT prediction and is comparable in magnitude to the Kondo temperature $T_K$. The match between experimental and DFT simulations confirms site-specific engineering of the diradical system.

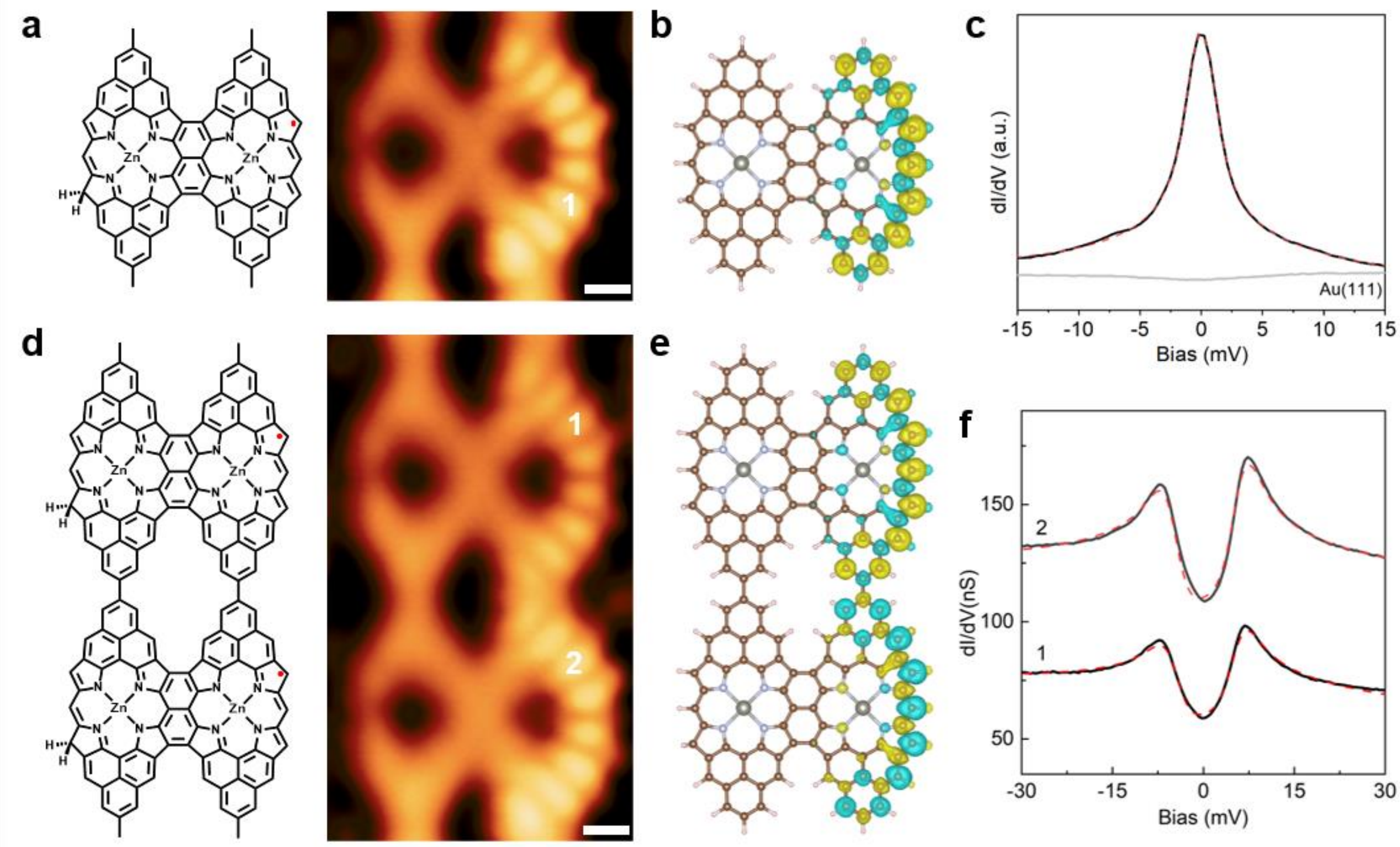

**Figure 2.** Magnetic properties of fused-double-porphyrin monomer and dimer. (a,d) Chemical structure and STM images (*V* =0.03 V, *I* =50 pA, Scale bar: 0.3 nm). (b,e) Spin density distribution calculated by DFT (yellow, spin up; blue, spin down). (c) d*I*/d*V* spectroscopy of fused-double-porphyrin monomer, showing a zero-bias Kondo resonance peak. The dashed line represents the d*I*/d*V* spectrum fitted with the Frota function. (f) d*I*/d*V* spectroscopy of fused-double-porphyrin dimer, showing two asymmetric low-bias peaks. The dashed line represents the d*I*/d*V* spectrum fitted with the Ternes model. Fitting yields antiferromagnetic exchange strength of 4.2 meV. The spectral acquisition points for (c) and (f) are marked on the STM images in (a) and (d), respectively.

We precisely constructed S=1/2 Kondo chains (up to 7 units) within fused-double-porphyrin frameworks. Systematic d*I*/d*V* measurements across the low-bias regime at each unit track the evolution of quantum states with chain length. As shown in Figures 3a and 4a, as the number of spin units increases, the range of the local density of states within the chain gradually expands. Moreover, increasing spin units generate significant site-dependence and parity effects: In odd-numbered chains (Figure 3b), zero-bias Kondo resonances appear at terminal/odd sites, with central sites exhibiting higher peak intensities than terminals, showing enhanced intensity; Even-numbered chains display asymmetric peaks near Fermi level, featuring narrower linewidths at terminals (Figure 4b). This parity-dependent evolution demonstrates that spin screening originates from emergent inter-site correlations rather than superposition of single-impurity Kondo physics.

To further interpret the observed parity-dependent spectral features, we adopt a complementary perspective on the spin-spin interactions used above. While the above fitting of $J_H$ attributed the gap in the diradical d*I*/d*V* spectrum to antiferromagnetic exchange interaction, we now consider an alternative framework in which each spin-1/2 radical forms a Kondo singlet with the metallic substrate, and the direct coupling appears (Figure. 3d). In this picture, wavefunction overlap between adjacent Kondo states gives rise to bonding and antibonding combinations,[12,52] which likewise produces splitting in the d*I*/d*V* spectrum. These two pictures—Heisenberg exchange and coupling Kondo chain—are often viewed as competing: strong antiferromagnetic coupling ($J_H \gg T_K$) favors an antiferromagnetic ground state and suppresses Kondo screening, whereas strong Kondo coupling ($J_H \ll T_K$) leads to fully screened local moments and quenched exchange[4]. However, when $J_H \sim T_K$, both mechanisms contribute comparably, and the two descriptions become effectively equivalent at low energies, which is the case here. We further discuss the connection between the two pictures in Supporting Information.

Motivated by this equivalence, we adopt a low-energy effective Hamiltonian that captures the coherent tunneling between screened Kondo singlets, with results shown in Figures 3c and 4c. At low temperatures, an individual radical coupled to the metal surface forms a Kondo singlet, manifesting as a resonance peak centered near the Fermi level with half-width $k_BT_K$. Introducing a second radical enables wavefunction overlap between adjacent Kondo singlets (Figure. 3d), permitting coherent electron tunneling that establishes direct coupling. We therefore employ the following low-energy effective Hamiltonian to simulate the system's density of states:

$$H_{\mathrm{eff}} = \sum_{n=1}^{N} \varepsilon_n a_n^\dagger a_n + \sum_{n=1}^{N-1} \left(t_{n,n+1} a_n^\dagger a_{n+1} + h.c.\right) \tag{1}$$

where $N$ is the number of radicals. $a_n$ is the annihilation operator describing the low-energy excitation of the $n$th Kondo singlet. The on-site energies $\varepsilon_n \approx 0$ represent the center of each individual Kondo resonance, while $t_{n,n+1}$ denotes the coupling between neighboring Kondo singlets. We compute the local density of states for this effective model (details in Supporting Information).

For N=2, the system can be effectively modeled by a two-site Hamiltonian. Simulations reveal the disappearance of the zero-bias peak, which splits into two peaks above and below the Fermi level—consistent with experimental diradical measurements (Figure S8). This behavior parallels bonding/antibonding state formation in coupled two-site systems.[12,52]

For longer Kondo chains, our model accurately captures the experimentally observed parity-dependent zero-energy states, which can be understood through the chiral symmetry of the effective Hamiltonian $H_{eff}$ satisfying $\{H_{\mathrm{eff}}, \hat{\Gamma}\} = 0$ with $\hat{\Gamma} = \sum_{n=1}^{N} (-1)^n a_n^\dagger a_n$. This symmetry divides the system into two sublattices: A-sites (odd positions, $N_A$) and B-sites (even positions, $N_B$). Choosing a basis ordered as all A-sublattice sites followed by all B-sublattice sites $\Psi^\dagger = (a_1^\dagger, a_3^\dagger, \ldots, a_{2N_A-1}^\dagger, a_2^\dagger, a_4^\dagger, \ldots, a_{2N_B}^\dagger)$, the chiral symmetry ensures that $H_{eff}$ can be represented in an off-diagonal block form:

$$H_{\mathrm{eff}} = \begin{pmatrix} 0 & T \\ T^\dagger & 0 \end{pmatrix} \tag{2}$$

where $T$ is an $N_A \times N_B$ matrix describing the hopping between the two sublattices. Because there is no hopping within the same sublattice, the diagonal blocks are zero. When the numbers of sites in the two sublattices differ, the rank of $T$ is generally $\min(N_A, N_B)$. This necessarily implies the existence of $|N_A - N_B|$ zero singular values of $T$, which correspond to zero-energy eigenstates of $H_{\text{eff}}$.[53] These zero-energy solutions reside entirely in the null space of either $T$ or $T^\dagger$, depending on which sublattice has more sites. As a result, the wavefunctions of the zero modes are completely supported on the sublattice with more sites, leading to exact sublattice polarization protected by chiral symmetry. Specifically, for odd-length chains ($N$ is odd, $N_A - N_B = 1$), sublattice imbalance generates unpaired zero-modes strictly localized at odd-numbered sites. Concurrently, wavefunction distribution in open-boundary chains enriches zero-modes at chain centers (Figure 3c), perfectly matching the experimental "center > terminal" zero-bias peak intensity hierarchy. For even-length chains ($N$ is even, $N_A - N_B = 0$), equal sublattice occupancy annihilates zero-modes, opening energy gaps in all local density of states and d*I*/d*V* spectra (Figure 4c). Therefore, the zero-energy modes of $H_{\text{eff}}$ are completely determined by the parity of the number of sites and the chiral symmetry. This parity-selective zero-mode behavior in the experiment serves as evidence for the direct coupling between Kondo singlets forming a coherent Kondo chain. Furthermore, in the Supporting Information, we include comparative calculations for weakly Kondo-screened and non-Kondo-screened S = 1/2 spin chains. These models fail to reproduce the experimental observations, further validating the key role of Kondo physics in our study.

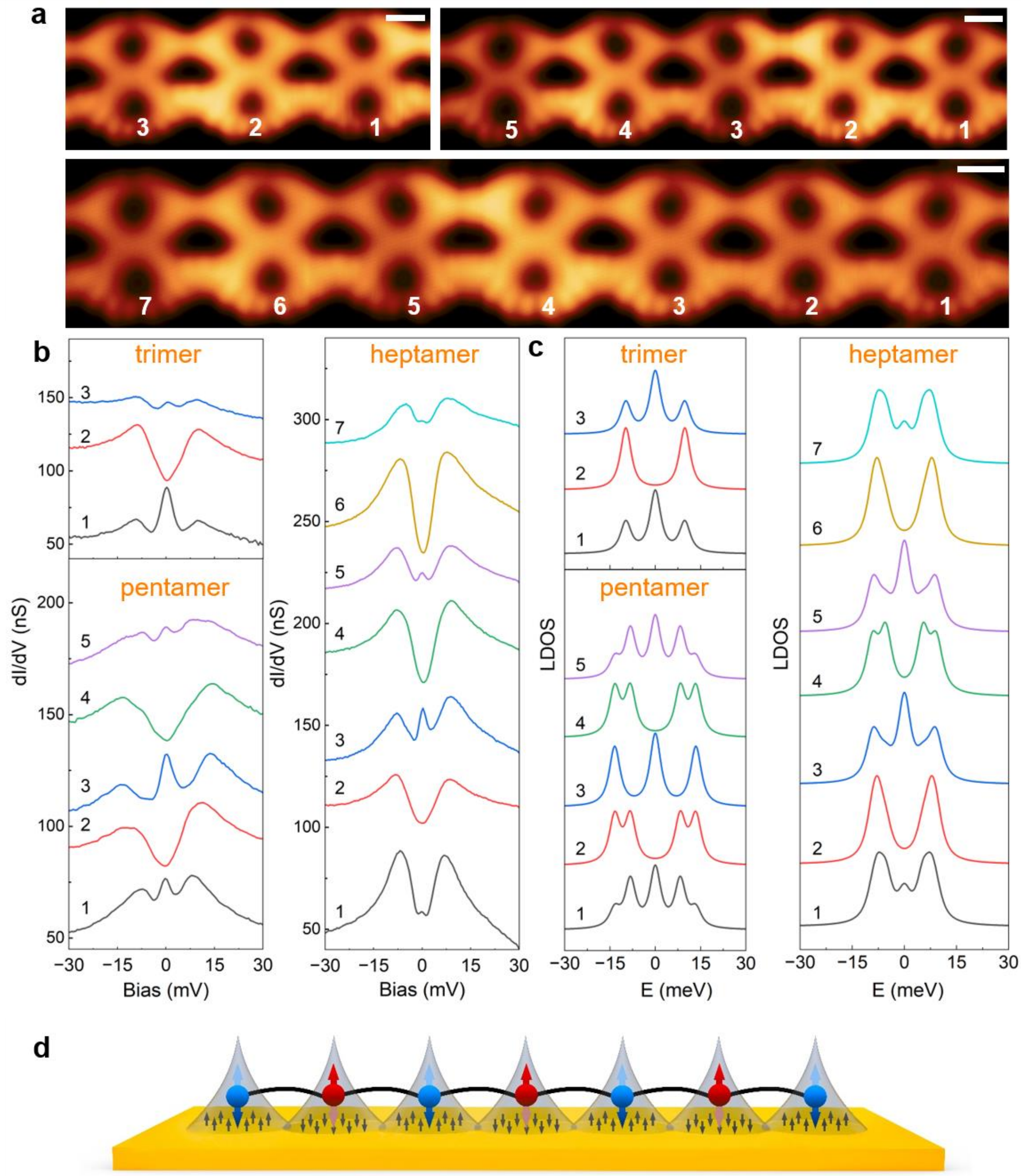

**Figure 3.** Stepwise construction and characterization of spin states in odd-numbered porphyrin chains. (a) Constant-current STM images of the selective formation of finite correlated Kondo chains via controlled tip-induced dehydrogenation ($V$ = 0.05 V, $I$ = 50 pA, Scale bar: 0.5 nm). (b) Corresponding d$I$/d$V$ spectra measured on the (a) chains. (c) Simulated d$I$/d$V$ spectra with 4 K thermal broadening. (d) Schematic diagram of the Kondo chain. The quantum spin states π-radical of individual porphyrin units are represented by bidirectional arrows, reflecting their isotropic S = 1/2 character without a preferred quantization axis. Adjacent singlets achieving coupling are denoted by gray shading.

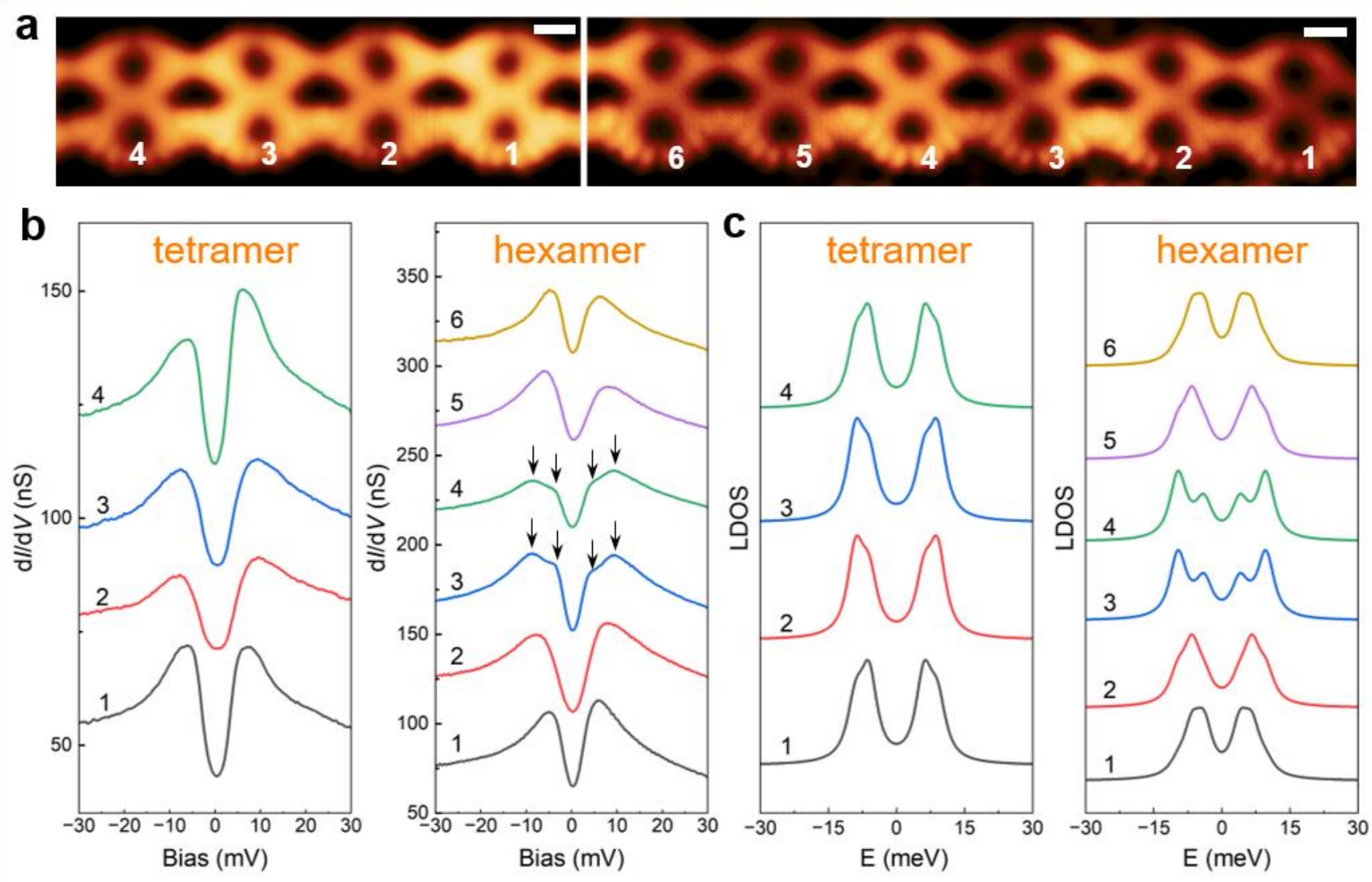

**Figure 4.** Stepwise construction and characterization of spin states in even-numbered porphyrin chains. (a) Constant-current STM images of the selective formation of finite correlated Kondo spin chains via stepwise tip-induced dehydrogenation ($V$ = 0.05 V, $I$ = 50 pA, Scale bar: 0.5 nm). (b) Corresponding d$I$/d$V$ spectra measured on the (a) chains. (c) Simulated d$I$/d$V$ spectra with 4 K thermal broadening.

Further investigation was conducted on how the number of parallel chains (single, fused-double, and fused-triple chains) differentially regulates the competition between the Kondo effect and spin exchange interaction. Initially, single spins are engineered on the same side of single and fused-triple-porphyrin chains. In both systems, clear Kondo resonances characteristic of a 1/2 spin state are observed. The Frota lineshape fitting to these spectra (Figure S9) yielded a HWHM of 1.42 meV and 1.38 meV, respectively. Subsequently, two spins were constructed on the same side, with STM measurements (Figure 5a,5d) and gas-phase DFT calculations confirming an antiferromagnetic spin ordering (Figure 5b,5e). The d$I$/d$V$ spectra for both single and fused-triple-porphyrin diradical exhibited asymmetric double peaks flanking the Fermi level (Figure 5c,5f). Fitting of multiple spectra using Ternes theory[52] revealed an exchange coupling strength $J_H$ in the range of 1.2–2.5 meV for the single-porphyrin dimers (Figure 5c and Figure S13b), which is significantly weaker than the value of 3.1–5.8 meV observed in fused-double-porphyrin dimers (Figure 2f and Figure S13c). Similarly, fused-triple-porphyrin dimers showed a reduced coupling of 1.5–3.2 meV (Figure 5f and Figure S13d) compared to the fused-double-porphyrin dimers. The difference in $J_H$ across these dimer systems arises from the combined effects of two distinct mechanisms. First, the enhanced coupling in fused-double-porphyrin dimers compared to single-porphyrin dimers stems from more efficient electron delocalization and thereby strengthens the wavefunction overlap between radical coupling sites (Table S2). Second, although the gas-phase computed $J_H$ value of the fused-triple-porphyrin is similar to that of the fused-double-porphyrin dimer, its experimentally observed reduction in in $J_H$ (2.8 meV vs. 4.2 meV) originates from a stronger Kondo screening effect. This system exhibits a narrower SOMO–LUMO gap (1.54 eV vs. 1.63 eV), which enhances hybridization with the substrate and increases the Kondo temperature $T_K$,[4] resulting in a shorter Kondo screening length $\xi_K$.

The change in $\xi_K$ can be further understood through the relation $\xi_K = \frac{\hbar v_F}{k_B T_K}$.[54] $\xi_K$ is inversely proportional to $T_K$, so an increase in $T_K$ leads to a reduction in $\xi_K$. A shorter $\xi_K$ indicates a more spatially confined screening cloud around each local moment, which weakens the overlap between adjacent Kondo states and thereby reducing the direct coupling term *t* in $H_{eff}$. This weakened coupling results in diminished level splitting, corresponding to a decrease in the effective spin exchange coupling strength. Consequently, fused-double-porphyrin chains exhibit stronger spin exchange than their fused-triple-chain counterparts. Within the $J_H \sim T_K$ regime, the competition exhibits a clear yet non-monotonic dependence on the number of parallel chains: Kondo interaction strength is strongest in the single chain, intermediate in the fused-triple-porphyrin chain, and weakest in the fused-double chain. Conversely, the spin exchange coupling follows an opposing trend: it is weakest in the single chain, intermediate in the fused-triple-porphyrin chain, and strongest in the fused-double chain. Finally, systematic studies of chain-length dependence (extending to six units, Figures S10–S11) revealed parity effects analogous to those in fused-double chains, demonstrating how finite-size effects fundamentally reconfigure many-body interactions in these confined quantum systems.

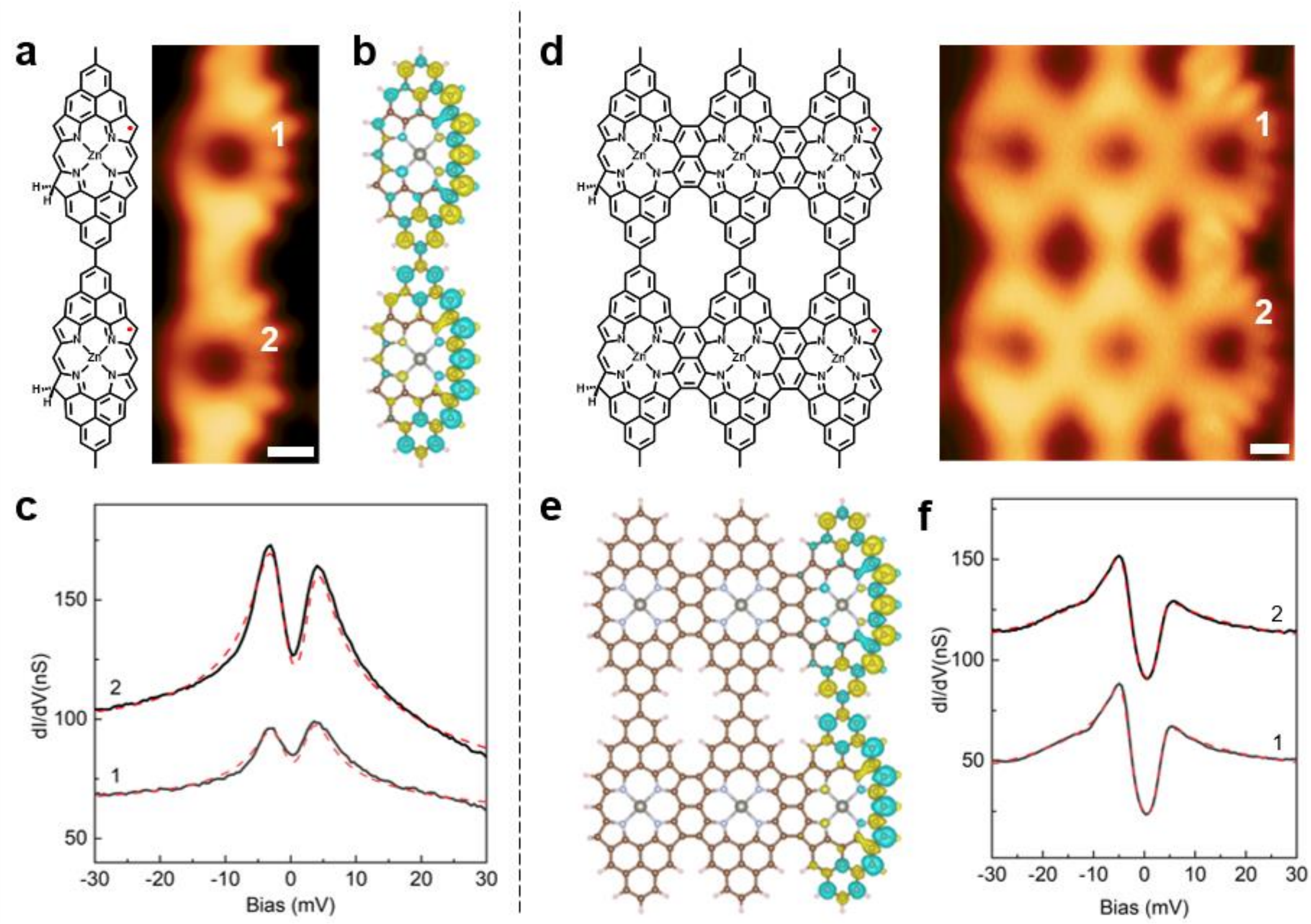

**Figure 5.** Width-dependent regulation of Kondo-magnetic exchange competition. (a,d) Chemical structures and corresponding constant-current STM images (*V* = 0.03 V, *I* = 30 pA, Scale bar: 0.3 nm ), (b,e) Spin density distribution calculated by DFT (yellow, spin up; blue, spin down), and (c,f) The solid

lines represent d*I*/d*V* spectroscopy of dimer configurations for both single porphyrin chains and fused-triple-porphyrin chains with two π-radicals, while the dashed lines show the corresponding fits based on the Ternes model. The fitting yields antiferromagnetic exchange strengths of 2.2 meV and 2.8 meV, respectively.

## Conclusions

This study employs a synergistic on-surface synthesis strategy utilizing C-C coupling of brominated $Zn^{2+}$-chelated porphyrin precursor (Zn(II)Por(dmp)$_2$-2Br), integrated with STM tip-selective *sp³* C-H bond dissociation. This approach achieves atomically precise construction of Kondo chains with π-radicals on Au(111). High-resolution STS reveals chain-length-dependent odd-even effects: odd chains exhibit centrally enhanced zero-bias peaks at odd sites (center > terminal), while even chains show exclusive asymmetric splitting near Fermi level. Simulations attribute this to chiral symmetry-induced sublattice partitioning, where odd chains develop unpaired zero-modes localized at odd sites (wavefunction concentrated at centers via open-boundary effects), and even chains exhibit gap formation from site-number parity. This theory-experiment match confirms coherent Kondo-singlet coupling. Furthermore, the number of parallel porphyrin chains—single, fused-double, or fused-triple—non-monotonically regulates the competition between the Kondo effect and spin exchange. The Kondo strength decreases as single > fused-triple > fused-double, while spin exchange follows the reverse trend. This behavior arises from two mechanisms: fused-double chains promote electron delocalization and enhance spin radical coupling, whereas the narrower SOMO–LUMO gap in fused-triple chains strengthens substrate hybridization, increasing $T_K$ and reducing spin coupling. This work not only establishes a new paradigm for precise control of strong correlations in confined quantum systems but also lays the experimental foundation for future quantum manipulation platforms based on molecular spin devices.

## Methods

**Sample preparation and STM/AFM experiments.** Sample preparation and STM/AFM experiments were conducted under ultrahigh vacuum (UHV) conditions. A Au(111) single crystal substrate was cleaned through repeated cycles of argon ion sputtering (1 keV, 15 minutes) followed by annealing at 773 K for 10 min, which yielded large, atomically flat terraces as verified by STM. The Zn(II)Por(dmp)$_2$-2Br precursors were synthesized specifically for this study, with the detailed procedure provided in the Supporting Information. These precursors were then deposited onto the clean surface at room temperature using a custom thermal evaporation source. A stepwise annealing protocol was employed to drive the on-surface reactions: annealing to 433 K induced debromination and covalent coupling, while subsequent annealing to 533 K resulted in near-complete cyclodehydrogenation, leading to the fusion of molecules into porphyrin-based oligomer chains with varying widths and lengths (Figure S2). To enhance spatial resolution for STM and non-contact AFM (nc-AFM) imaging, the sample was cooled to approximately 10 K and functionalized by transferring a single CO molecule from the surface to the apex of an electrochemically etched tungsten tip.

All STM topographic images were acquired in constant-current mode, and scanning tunneling spectroscopy (STS) measurements were performed using a lock-in amplifier with a modulation voltage of 1 mV at 983 Hz. Nc-AFM imaging were carried out using a qPlus sensor (resonance frequency ~32 kHz) in constant-height mode with a minimal sample bias of 1 mV, with all spectroscopic data collected at 4.2 K.

**DFT calculations.** Spin-polarized DFT calculations were carried out using the Gaussian 16 software package. The PBE0 functiona was employed to describe the electronic structure of all free-standing radicals. Geometry optimization was performed with the def2-SVP basis set, with the dihedral angle at the radical coupling site constrained to 180° to mimic the planar configuration after surface adsorption. Subsequent single-point energy calculations employed the def2-TZVP basis set to improve energetic accuracy. Molecular structures and orbital isosurfaces were visualized using VESTA.

## Acknowledgements

This work is supported by the National Key R&D Program of China (2024YFA1208202, 2024YFA1409002), the National Natural Science Foundation of China (22225202, T2522001, 92477207, 92356309, 22132007, 22172002, T2588301 22402004, 22302068, 52173205, 12374034), the Shanghai Sailing Program (23YF1408700), and the Innovation Program for Quantum Science and Technology (2021ZD0302403). DFT calculations are carried out on TianHe-1A at National Supercomputer Center in Tianjin and supported by High-performance Computing Platform of Peking University. Experiments are supported by Peking Nanofab.

## ASSOCIATED CONTENT

### Supporting Information

Detailed descriptions of precursor/on-surface synthesis, STM manipulation, theoretical computations (DFT, numerical simulation protocols, 1/2 spin chains, thermal broadening), expanded STM/Nc-AFM characterization, STS/simulated STS analyses, and supplementary results (PDF).

### Author Contributions

†These authors contributed equally.

### Notes

The authors declare no competing financial interests.

### Data availability.

The datasets generated and/or analysed during the current study are available from the corresponding author upon reasonable request.

**References**

(1) Auerbach, A. *Interacting Electrons and Quantum Magnetism*; Birman, J. L., Lynn, J. W., Silverman, M. P., Stanley, H. E., Voloshin, M., Series Eds.; Graduate Texts in Contemporary Physics; Springer New York: New York, NY, **1994**.

(2) Paschen, S.; Si, Q. Quantum Phases Driven by Strong Correlations. *Nat. Rev. Phys.* **2020**, *3* (1), 9–26.

(3) Dagotto, E. Complexity in Strongly Correlated Electronic Systems. *Science* **2005**, *309* (5732), 257–262.

(4) Hewson, A. C. The Kondo Problem to Heavy Fermions, 1st ed.; *Cambridge University Press*, **1993**.

(5) Si, Q. Quantum Criticality and Global Phase Diagram of Magnetic Heavy Fermions. *Phys. Status Solidi B* **2010**, *247* (3), 476–484.

(6) Li, J.; Schneider, W.-D.; Berndt, R.; Delley, B. Kondo Scattering Observed at a Single Magnetic Impurity. *Phys. Rev. Lett.* **1998**, *80* (13), 2893–2896.

(7) Crommie, M. F. Manipulating Magnetism in a Single Molecule. *Science* **2005**, *309* (5740), 1501–1502.

(8) Choi, D.-J.; Lorente, N.; Wiebe, J.; Von Bergmann, K.; Otte, A. F.; Heinrich, A. J. *Colloquium*: Atomic Spin Chains on Surfaces. *Rev. Mod. Phys.* **2019**, *91* (4), 041001.

(9) Cronenwett, S. M.; Oosterkamp, T. H.; Kouwenhoven, L. P. A Tunable Kondo Effect in Quantum Dots. *Science* **1998**, *281* (5376), 540–544.

(10) Madhavan, V.; Chen, W.; Jamneala, T.; Crommie, M. F.; Wingreen, N. S. Tunneling into a Single Magnetic Atom: Spectroscopic Evidence of the Kondo Resonance. *Science* **1998**, *280* (5363), 567–569.

(11) Heinrich, A. J.; Gupta, J. A.; Lutz, C. P.; Eigler, D. M. Single-Atom Spin-Flip Spectroscopy. *Science* **2004**, *306* (5695), 466–469.

(12) Jeong, H.; Chang, A. M.; Melloch, M. R. The Kondo Effect in an Artificial Quantum Dot Molecule. *Science* **2001**, *293* (5538), 2221–2223.

(13) Tsukahara, N.; Shiraki, S.; Itou, S.; Ohta, N.; Takagi, N.; Kawai, M. Evolution of Kondo Resonance from a Single Impurity Molecule to the Two-Dimensional Lattice. *Phys. Rev. Lett.* **2011**, *106* (18), 187201.

(14) Jiang, Y.; Zhang, Y. N.; Cao, J. X.; Wu, R. Q.; Ho, W. Real-Space Imaging of Kondo Screening in a Two-Dimensional $O_2$ Lattice. *Science* **2011**, *333* (6040), 324–328.

(15) Wan, W.; Harsh, R.; Meninno, A.; Dreher, P.; Sajan, S.; Guo, H.; Errea, I.; De Juan, F.; Ugeda, M. M. Evidence for Ground State Coherence in a Two-Dimensional Kondo Lattice. *Nat. Commun.* **2023**, *14* (1), 7005.

(16) García-Martínez, N. A., Lado, J. L., Jacob, D., Fernández-Rossier, J. Anomalous magnetism in hydrogenated graphene. *Phys. Rev. B* **2017**, *96*(2), 024403.

(17) Lieb, E. H. Two Theorems on the Hubbard Model. *Phys. Rev. Lett.* **1989**, *62* (10), 1201–1204.

(18) Ovchinnikov, A. A. Multiplicity of the Ground State of Large Alternant Organic Molecules with Conjugated Bonds: (Do Organic Ferromagnetics Exist?). *Theor. Chim. Acta* **1978**, *47* (4), 297–304.

(19) Mishra, S.; Beyer, D.; Eimre, K.; Liu, J.; Berger, R.; Gröning, O.; Pignedoli, C. A.; Müllen, K.; Fasel, R.; Feng, X.; Ruffieux, P. Synthesis and Characterization of π-Extended Triangulene. *J. Am. Chem. Soc.* **2019**, *141* (27), 10621–10625.

(20) Mishra, S.; Beyer, D.; Eimre, K.; Kezilebieke, S.; Berger, R.; Gröning, O.; Pignedoli, C. A.; Müllen, K.; Liljeroth, P.; Ruffieux, P.; Feng, X.; Fasel, R. Topological Frustration Induces Unconventional Magnetism in a Nanographene. *Nat. Nanotechnol.* **2020**, *15* (1), 22–28.

(21) Li, J.; Sanz, S.; Castro-Esteban, J.; Vilas-Varela, M.; Friedrich, N.; Frederiksen, T.; Peña, D.; Pascual, J. I. Uncovering the Triplet Ground State of Triangular Graphene Nanoflakes Engineered with Atomic Precision on a Metal Surface. *Phys. Rev. Lett.* **2020**, *124* (17), 177201.

(22) Su, J.; Lyu, P.; Lu, J. Atomically Precise Imprinting π-Magnetism in Nanographenes via Probe Chemistry. *Precis. Chem.* **2023**, *1* (10), 565–575.

(23) Vegliante, A.; Vilas-Varela, M.; Ortiz, R.; Romero Lara, F.; Kumar, M.; Gómez-Rodrigo, L.; Trivini, S.; Schulz, F.; Soler-Polo, D.; Ahmoum, H.; Artacho, E.; Frederiksen, T.; Jelínek, P.; Pascual, J. I.; Peña, D. On-Surface Synthesis of a Ferromagnetic Molecular Spin Trimer. *J. Am. Chem. Soc.* **2025**, jacs.4c15736.

(24) Min, H.; Hill, J. E.; Sinitsyn, N. A.; Sahu, B. R.; Kleinman, L.; MacDonald, A. H. Intrinsic and Rashba Spin-Orbit Interactions in Graphene Sheets. *Phys. Rev. B* **2006**, *74* (16), 165310.

(25) Yazyev, O. V. Hyperfine Interactions in Graphene and Related Carbon Nanostructures. *Nano Lett.* **2008**, *8* (4), 1011–1015.

(26) Lombardi, F.; Lodi, A.; Ma, J.; Liu, J.; Slota, M.; Narita, A.; Myers, W. K.; Müllen, K.; Feng, X.; Bogani, L. Quantum Units from the Topological Engineering of Molecular Graphenoids. *Science* **2019**, *366* (6469), 1107–1110.

(27) Clair, S.; De Oteyza, D. G. Controlling a Chemical Coupling Reaction on a Surface: Tools and Strategies for On-Surface Synthesis. *Chem. Rev.* **2019**, *119* (7), 4717–4776.

(28) De Oteyza, D. G.; Frederiksen, T. Carbon-Based Nanostructures as a Versatile Platform for Tunable π-Magnetism. *J. Phys. Condens. Matter* **2022**, *34* (44), 443001.

(29) Mishra, S.; Catarina, G.; Wu, F.; Ortiz, R.; Jacob, D.; Eimre, K.; Ma, J.; Pignedoli, C. A.; Feng, X.; Ruffieux, P.; Fernández-Rossier, J.; Fasel, R. Observation of Fractional Edge Excitations in Nanographene Spin Chains. *Nature* **2021**, *598* (7880), 287–292.

(30) Zhao, Y.; Jiang, K.; Li, C.; Liu, Y.; Zhu, G.; Pizzochero, M.; Kaxiras, E.; Guan, D.; Li, Y.; Zheng, H.; Liu, C.; Jia, J.; Qin, M.; Zhuang, X.; Wang, S. Quantum Nanomagnets in On-Surface Metal-Free Porphyrin Chains. *Nat. Chem.* **2023**, *15* (1), 53–60.

(31) Zhao, C.; Catarina, G.; Zhang, J.-J.; Henriques, J. C. G.; Yang, L.; Ma, J.; Feng, X.; Gröning, O.; Ruffieux, P.; Fernández-Rossier, J.; Fasel, R. Tunable Topological Phases in Nanographene-Based Spin-1/2 Alternating-Exchange Heisenberg Chains. *Nat. Nanotechnol.* **2024**, *19* (12), 1789–1795.

(32) Yuan, Z.; Zhang, X.-Y.; Jiang, Y.; Qian, X.; Wang, Y.; Liu, Y.; Liu, L.; Liu, X.; Guan, D.; Li, Y.; Zheng, H.; Liu, C.; Jia, J.; Qin, M.; Liu, P.-N.; Li, D.-Y.; Wang, S. Fractional Spinon Quasiparticles in Open-Shell Triangulene Spin-1/2 Chains. *J. Am. Chem. Soc.* **2025**, *147* (6), 5004–5013.

(33) Sun, K.; Cao, N.; Silveira, O. J.; Fumega, A. O.; Hanindita, F.; Ito, S.; Lado, J. L.; Liljeroth, P.; Foster, A. S.; Kawai, S. On-Surface Synthesis of Heisenberg Spin-1/2 Antiferromagnetic Molecular Chains. *Sci. Adv.* **2025**, *11* (9), eads1641.

(34) Su, X.; Ding, Z.; Hong, Y.; Ke, N.; Yan, K.; Li, C.; Jiang, Y.-F.; Yu, P. Fabrication of Spin-1/2 Heisenberg Antiferromagnetic Chains via Combined on-Surface Synthesis and Reduction for Spinon Detection. *Nat. Synth.* **2025**. *4* (6), 694-701.

(35) Fu, X.; Huang, L.; Liu, K.; Henriques, J. C. G.; Gao, Y.; Han, X.; Chen, H.; Wang, Y.; Palma, C.-A.; Cheng, Z.; Lin, X.; Du, S.; Ma, J.; Fernández-Rossier, J.; Feng, X.; Gao, H.-J. Building Spin-1/2 Antiferromagnetic Heisenberg Chains with Diaza-Nanographenes. *Nat. Synth.* **2025**. *4* (6), 684-693.

(36) Zhao, C.; Yang, L.; Henriques, J. C. G.; Ferri-Cortés, M.; Catarina, G.; Pignedoli, C. A.; Ma, J.; Feng, X.; Ruffieux, P.; Fernández-Rossier, J.; Fasel, R. Spin Excitations in Nanographene-Based Antiferromagnetic Spin-1/2 Heisenberg Chains. *Nat. Mater.* **2025**, *24* (5), 722–727.

(37) Choi, D.-J.; Robles, R.; Yan, S.; Burgess, J. A. J.; Rolf-Pissarczyk, S.; Gauyacq, J.-P.; Lorente, N.; Ternes, M.; Loth, S. Building Complex Kondo Impurities by Manipulating Entangled Spin Chains. *Nano Lett.* **2017**, *17* (10), 6203–6209.

(38) Moro-Lagares, M.; Korytár, R.; Piantek, M.; Robles, R.; Lorente, N.; Pascual, J. I.; Ibarra, M. R.; Serrate, D. Real Space Manifestations of Coherent Screening in Atomic Scale Kondo Lattices. *Nat. Commun.* **2019**, *10* (1), 2211.

(39) DiLullo, A.; Chang, S.-H.; Baadji, N.; Clark, K.; Klöckner, J.-P.; Prosenc, M.-H.; Sanvito, S.; Wiesendanger, R.; Hoffmann, G.; Hla, S.-W. Molecular Kondo Chain. *Nano Lett.* **2012**, *12* (6), 3174–3179.

(40) Wen, E. C. H.; Jacobse, P. H.; Jiang, J.; Wang, Z.; McCurdy, R. D.; Louie, S. G.; Crommie, M. F.; Fischer, F. R. Magnetic Interactions in Substitutional Core-Doped Graphene Nanoribbons. *J. Am. Chem. Soc.* **2022**, *144* (30), 13696–13703.

(41) Sun, Q.; Mateo, L. M.; Robles, R.; Lorente, N.; Ruffieux, P.; Bottari, G.; Torres, T.; Fasel, R. Bottom-up Fabrication and Atomic-Scale Characterization of Triply Linked, Laterally π-Extended Porphyrin Nanotapes. *Angew. Chem. Int. Ed.* **2021**, *60* (29), 16208–16214.

(42) Wang, T.; Berdonces-Layunta, A.; Friedrich, N.; Vilas-Varela, M.; Calupitan, J. P.; Pascual, J. I.; Peña, D.; Casanova, D.; Corso, M.; De Oteyza, D. G. Aza-Triangulene: On-Surface Synthesis and Electronic and Magnetic Properties. *J. Am. Chem. Soc.* **2022**, *144* (10), 4522–4529.

(43) Ternes, M.; Heinrich, A. J.; Schneider, W.-D. Spectroscopic Manifestations of the Kondo Effect on Single Adatoms. *J. Phys. Condens. Matter* **2009**, *21* (5), 053001.

(44) Zheng, Y.; Li, C.; Zhao, Y.; Beyer, D.; Wang, G.; Xu, C.; Yue, X.; Chen, Y.; Guan, D.-D.; Li, Y.-Y.; Zheng, H.; Liu, C.; Luo, W.; Feng, X.; Wang, S.; Jia, J. Engineering of Magnetic Coupling in Nanographene. *Phys. Rev. Lett.* **2020**, *124* (14), 147206.

(45) Li, J.; Sanz, S.; Corso, M.; Choi, D. J.; Peña, D.; Frederiksen, T.; Pascual, J. I. Single Spin Localization and Manipulation in Graphene Open-Shell Nanostructures. *Nat. Commun.* **2019**, *10* (1), 200.

(46) Zhao, Y.; Jiang, K.; Li, C.; Liu, Y.; Xu, C.; Zheng, W.; Guan, D.; Li, Y.; Zheng, H.; Liu, C.; Luo, W.; Jia, J.; Zhuang, X.; Wang, S. Precise Control of π-Electron Magnetism in Metal-Free Porphyrins. *J. Am. Chem. Soc.* **2020**, *142* (43), 18532–18540.

(47) Sun, Q.; Mateo, L. M.; Robles, R.; Ruffieux, P.; Lorente, N.; Bottari, G.; Torres, T.; Fasel, R. Inducing Open-Shell Character in Porphyrins through Surface-Assisted Phenalenyl π-Extension. *J. Am. Chem. Soc.* **2020**, *142* (42), 18109–18117.

(48) Turco, E.; Aapro, M.; Ganguli, S. C.; Krane, N.; Drost, R.; Sobrino, N.; Bernhardt, A.; Juríček, M.; Fasel, R.; Ruffieux, P.; Liljeroth, P.; Jacob, D. Demonstrating Kondo Behavior by Temperature-Dependent Scanning Tunneling Spectroscopy. *Phys. Rev. Res.* **2024**, *6* (2), L022061.

(49) Frota, H. O. Shape of the Kondo Resonance. *Phys. Rev. B* **1992**, 45 (3), 1096–1099.

(50) Madhavan, V.; Chen, W.; Jamneala, T.; Crommie, M. F.; Wingreen, N. S. Local Spectroscopy of a Kondo Impurity: Co on Au(111). *Phys. Rev. B* **2001**, *64* (16), 165412.

(51) Ternes, M. Spin Excitations and Correlations in Scanning Tunneling Spectroscopy. *New J. Phys.* **2015**, *17* (6), 063016.

(52) Esat, T.; Lechtenberg, B.; Deilmann, T.; Wagner, C.; Krüger, P.; Temirov, R.; Rohlfing, M.; Anders, F. B.; Tautz, F. S. A Chemically Driven Quantum Phase Transition in a Two-Molecule Kondo System. *Nat. Phys.* **2016**, *12* (9), 867–873.

(53) Inui, M.; Trugman, S. A.; Abrahams, E. Unusual Properties of Midband States in Systems with Off-Diagonal Disorder. *Phys Rev B* **1994**, *49* (5), 3190–3196.

(54) V. Borzenets, I.; Shim, J.; Chen, J. C. H.; Ludwig, A.; Wieck, A. D.; Tarucha, S.; Sim, H.-S.; Yamamoto, M. Observation of the Kondo Screening Cloud. *Nature* **2020**, *579* (7798), 210–213.

**TOC graphic**

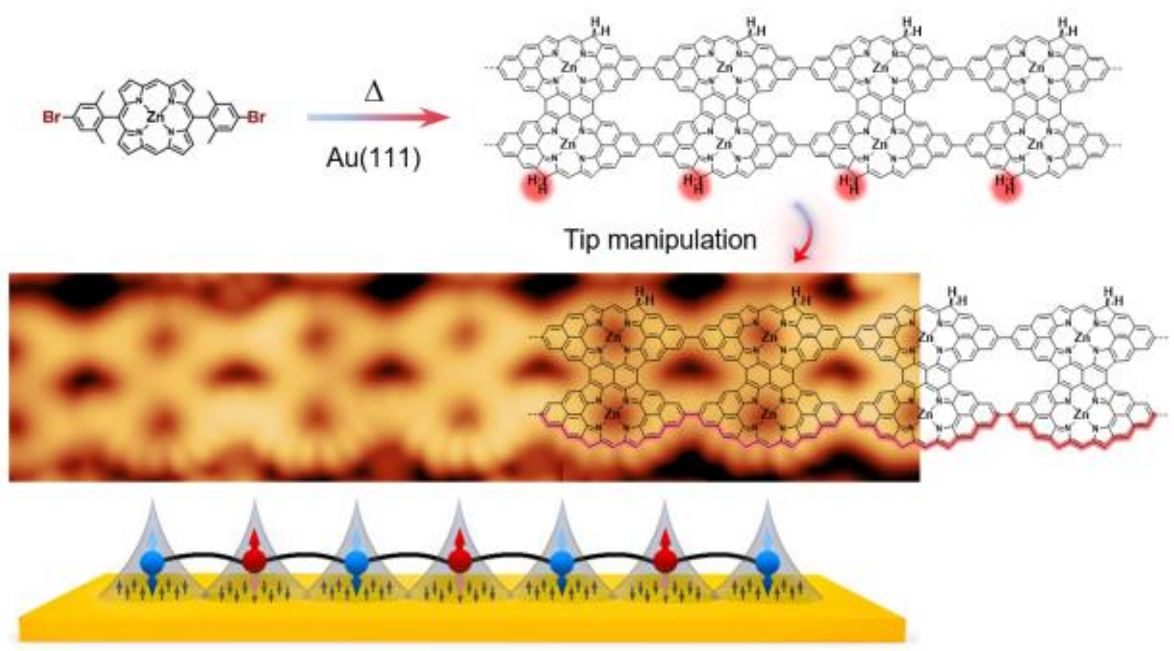